\documentclass[preprint,showkeys]{revtex4} 
\usepackage{amsmath}
\usepackage{latexsym}
\usepackage{amssymb}
\usepackage{graphics,epstopdf}
\usepackage{amsmath,amssymb,amsthm}
\usepackage{graphicx}
\usepackage[colorlinks=true, citecolor=blue, urlcolor=blue ]{hyperref}
\usepackage{cleveref}

\begin{document}
	\title{On the entanglement of co-ordinate and momentum degrees of freedom in noncommutative space}
	\author{Shilpa Nandi}
	\email{nandishilpa801325@gmail.com}
	\address{Dept. of Physics, Acharya Prafulla Chandra College, New Barrackpore, 24 pgs (N), West Bengal, India- 700131.}
	\author{Muklesur Rahaman}
	\email{mrrahaman360@gmail.com}
	\address{Dept. of Physics, Acharya Prafulla Chandra College, New Barrackpore, 24 pgs (N), West Bengal, India- 700131.}
	\author{Pinaki Patra}
	\thanks{Corresponding author}
	\email{monk.ju@gmail.com}
	\affiliation{Department of Physics, Brahmananda Keshab Chandra College, Kolkata, India-700108}
	
	
	\date{\today}

	\begin{abstract}
	In this paper, we investigate the quantum entanglement induced by phase-space noncommutativity. Both the position-position and momentum-momentum noncommutativity are incorporated to study the entanglement properties of coordinate and momentum degrees of freedom under the shade of oscillators in noncommutative space.  Exact solutions for the systems are obtained after the model is re-expressed in terms of canonical variables, by performing a particular Bopp's shift to the noncommuting degrees of freedom.
	It is shown that the bipartite Gaussian state for an isotropic oscillator is always separable. To extend our study for the time-dependent system, we allow arbitrary time dependency on parameters. The time-dependent isotropic oscillator is solved with the Lewis-Riesenfeld invariant method. It turns out that even for arbitrary time-dependent scenarios, the separability property does not alter.
	\\
	We extend our study to the anisotropic oscillator, which provides an entangled state even for time-independent parameters.  The Wigner quasi-probability distribution is constructed for a bipartite Gaussian state. The noise matrix (covariance matrix) is explicitly studied with the help of Wigner distribution.  Simon's separability criterion (generalized Peres-Horodecki criterion) has been employed to find the unique function of the (mass and frequency) parameters, for which the bipartite states are separable.  In particular, we show that the mere inclusion of non-commutativity of phase-space is not sufficient to generate the entanglement, rather anisotropy is important at the same footing.
	\end{abstract}
	
	\maketitle

\section{Introduction}
Quantum entanglement is not spooky-it is a real fact, it exists. Quantum entanglement has been already spotted in objects almost big enough to see. For instance,  a remote quantum entanglement between two micromechanical oscillators was reported in \cite{enatnglementexpt1}.
Quantum entanglement (verschr\"{a}nkung) implies the existence of global states of a composite system which can not be written as a product of the states of individual
subsystems \cite{entanglement1}. In other words,  entanglement underlines the intrinsic order of statistical relations between subsystems of a compound quantum system \cite{entanglement2}. We expect that the insight into quantum entanglement will lead to an understanding of a diverse range of phenomena, in which correlations are important. For instance, the understanding of quantum optics, phase transition in condensed matter systems, black hole physics, and the relation between the geometrical structure of the dual spacetime and entanglement structure of the conformal field theory make the quantum entanglement a topical issue \cite{entanglement3,entanglement4,entanglement5,entanglement6}. Nonetheless, the physical origin of quantum entanglement is still obscure. One of the key goals of quantum information theory is to understand the physical origin(s) of entanglement in quantum systems. 
Such a goal is unlikely to be completely realized unless the role of the noncommutativity on the underlying phase-space on the correlation between coordinate degrees of freedom is explored. 
\\
 It seems likely that any modification to the symplectic structure of phase-space \cite{symplectic1}, as brought about by position-position noncommutativity and momentum-momentum noncommutativity at a fundamental scale, could impact the correlations of co-ordinate and momentum degrees of freedom. We expect a deep-rooted connection between noncommutative dynamics and the phenomenon of entanglement of quantum states. We would like to mention that this modification is not merely just computational pleasure. Indeed, it is directly related to the quantum theory of gravity-most promising endeavor of modern physics. 
 It is almost a Gospel that the fundamental concept of space-time is mostly compatible with quantum theory in noncommutative space (NCS) \cite{ncs1}. Formalisms of Quantum theory in NCS have been applied to various physical situations \cite{ncsho,ncsosc,ncstopics,ncsfluid}. For the reviews on NCS, one can see \cite{ncsreview1,ncsreview2}.  
 \\
 Although there are several promising proposals for quantum gravity (QG),  the unified theory of QG lacks direct experimental evidence. Most probably, the lack of successful experiments for QG is not due to the drawback of the proposed theories. Rather, it is due to the limitation of present-day experimental capacity. By way of comparison, the LHC was designed to run at a maximum collision energy of $14$  TeV \cite{lhc}, whereas most of the theories of quantum gravity appear to predict departures from classical gravity only at energy scales on the order of $10^{19}$ GeV \cite{qg1,qg2}. We hope that the high-energy scattering experiment may lead us to the feasibility of probing such high-energy scales in the future. However, colliders are costly. Naturally, alternative directions toward the exploration of the quantum signature of gravity are always fascinating. For instance,  proposals with the help of conventional interferometric techniques are available in the literature \cite{expt1,expt2,expt3}. The optomechanical scheme relies on a massive noncommutative oscillator to interact with a strong optical field in an optomechanical cavity by utilizing the principle of radiation pressure interaction \cite{expt4,expt5}.   A deformation of the commutation relation will induce a measurable deviation in the optical phase with respect to that of the usual canonical commutation relation.
 From a viewpoint of constructor-theoretic principles, the proposed optomechanical schemes are promising. In particular, if we observe the entanglement effects
 in the measurement of the properties of two quantum masses that interact with each other
 through gravity only, then we can conclude that the mediator (gravity) has to have some
 quantum features \cite{marletto1}. It doesn’t matter in what way gravity is quantum - whether it is loop quantum gravity or string theory or something else - but it has to be a quantum theory \cite{pinaki1}. \\ 
 Therefore, the entanglement properties in NCS are worth studying.  We would like to emphasize that the topic is not new. Several studies in this regime have enriched the literature already. For instance,  the role of position-space noncommutativity on the entanglement of bipartite systems for  Gaussian states is reported in \cite{ncsentangled1}. The quantum nature and the separability criterion of noncommutative two-mode Gaussian states were illustrated by Bastos, et al. \cite{ncsentangled2}. Rendell et al. examined the entanglement of general asymmetric pure Gaussian two-mode state in terms of the coefficients of the quadrature components of the wave function \cite{ncsentangled3}. Quantum entanglement, induced by spatial noncommutativity,  for an anisotropic harmonic oscillator is explicitly described in \cite{ncsentangled4}.  Gaussian entanglement for an anisotropic harmonic oscillator with arbitrarily time-dependent parameters is also available in the literature \cite{ncsentangled5}.
\\
In the present study, we explore whether the incorporation of position-position, as well as momentum-momentum noncommutativity, are sufficient to generate entanglement for the coordinate degrees of freedom. In particular, we address the issue of entanglement induced by noncommutative (NC) parameters. NC parameters can indeed generate entanglement between coordinate degrees of freedom for pure Gaussian states. However, we show that the incorporation of NC parameters is not the whole story behind the Gaussian entanglement. In order to show the entanglement, the oscillators have to be anisotropic. Both the NC parameters and anisotropy in oscillators are necessary for entanglement in coordinate degrees of freedom. 
\\
In the present study, we restrict ourselves to the entanglement of
bipartite Gaussian states for which the standard symplectic
formalism is readily available and well established in the commutative limit. To make our study applicable to diverse domains, we consider the isotropic oscillators (IO) with time-dependent (TD) parameters. However, we restrict ourselves to the parameter values for which the Hamiltonian under consideration is self-adjoint under the usual inner product. With the help of the Lewis-Riesenfeld (LR) phase-space invariant method we first solve the TD-Hamiltonian for IO. 
Lewis-Riesenfeld (LR) invariant method is based on the construction of eigenstates of a TD- invariant operator $\hat{\mathcal{I}}(t)$ corresponding to the Hamiltonian $\hat{H}(t)$, and TD-phase factors from the consistency of the solution of TD-Schr\"{o}dinger equation of $\hat{H}(t)$  \cite{Lewis1,Lewis2}. The LR-invariant operator is transformed into a simpler form with $Sp(4,\mathbb{R})$ symplectic diagonalization. Through the eigenstates of the invariant operator, we show that the bipartite Gaussian states are always separable for isotropic cases. However, for an anisotropic oscillator, the states are entangled even for time-independent parameters. The anisotropic oscillator is exactly solved. Since the covariance matrix can be straightforwardly constructed with the help of the expectation values for Wigner quasiprobability distribution corresponding to the state of the system \cite{gaussian1,wigner1,wigner2},  we construct the covariance matrix ($\hat{\mathcal{V}}$) through the Wigner distribution corresponding to the Gaussian state of time-independent anisotropic oscillator. Gaussian states are well characterized by the bonafide covariance matrix. 
We then apply the generalized Peres-Horodecki separability criterion (Simon's condition), which is specified through $\hat{\mathcal{V}}$ \cite{simon1}.  Simon's condition is a sufficient condition for the separability of Gaussian states.  It is worth mentioning that, since the connection (Darboux transformation) between the commutative space and NC-space is not symplectic,   the separability criterion for NC-space is not immediate  \cite{rsupnc1,rsupnc2}.  However, one can transform the NC-space system to an equivalent system in commutative space through Bopp's shift (Darboux transformation) and use the formalism of usual commutative space quantum mechanics, including the Simon's separability condition \cite{rsupnc1,rsupnc2,rsupnc4}. This leads to some restrictions on the anisotropic parameters in order for the bipartite state to be separable.  \\
The organization of the present paper is as follows. First, we apply Bopp's shift to transform our NC-space oscillators into an equivalent commutative space oscillator. Then we construct Lewis-Riesenfeld invariant operators corresponding to the Hamiltonian of an isotropic oscillator in commutative space. The invariant operator is transformed into a simpler form with symplectic diagonalization (symplectic group $Sp(4,\mathbb{R})$). Eigenstates of the invariant operator, are constructed so that we have the solutions of TD-Schr\"{o}dinger equation corresponding to our system. We show that the bipartite Gaussian state for an isotropic oscillator is always separable, irrespective of the choice of TD parameters. Then we investigate the scenario for an anisotropic oscillator. The quantum noise matrix (covariance matrix) corresponding to the Wigner distribution for the Gaussian state is constructed. We then apply Simon's separability criterion to determine the constraint on anisotropic parameters to characterize the entanglement properties of coordinate degrees of freedom. Finally, we conclude our result with experimental implications.
\section{NC-space oscillator and corresponding commutative space system}
We use the notation
$\tilde{X}= (\hat{\tilde{X}}_1,\hat{\tilde{X}}_2, \hat{\tilde{X}}_3, \hat{\tilde{X}}_4)^T= (\hat{\tilde{x}}_1,\hat{\tilde{p}}_1, \hat{\tilde{x}}_2, \hat{\tilde{p}}_2)^T$, for  a coordinate vector in noncommutative (NC) space. The same for the usual commutative space is denoted by	$X=(\hat{X}_1, \hat{X}_2,\hat{X}_3,\hat{X}_4)^T=(\hat{x}_1, \hat{p}_1,\hat{x}_2,\hat{p}_2)^T$. Here $X^T$ stands for the matrix transposition of $X$.
The  commutation relations for   commutative space reads
\begin{equation}\label{Cspacecommutation}
	[	\hat{X}_{\alpha}, \hat{X}_{\beta}]=i\hbar \hat{J}_{\alpha\beta}= -\hbar(\hat{\Sigma}_y)_{\alpha\beta}, \;
\mbox{with}\;
	\hat{J}=\mbox{diag}(\hat{J}_2,\hat{J}_2),\;\; \hat{\Sigma}_y=\mbox{diag}(\hat{\sigma}_y, \hat{\sigma}_y).
\end{equation}
Here the symplectic matrix $\hat{J}_2$ and Pauli matrices are represented by
\begin{eqnarray}
	\hat{\sigma}_x =\left(\begin{array}{cc}
		0 & 1\\
		1& 0
	\end{array}\right),\;
	\hat{\sigma}_y =\left(\begin{array}{cc}
		0 & -i\\
		i& 0
	\end{array}\right),\;
	\hat{\sigma}_z =\left(\begin{array}{cc}
		1 & 0\\
		0& -1
	\end{array}\right),\;
\hat{J}_2 = \left(\begin{array}{cc}
	0 & 1\\
	-1& 0
\end{array}\right).
\end{eqnarray}
We adopt the fundamental commutation relations  in NC-space  as follows.
\begin{equation}\label{NCspacecommutation}
	[	\hat{\tilde{X}}_{\alpha}, \hat{\tilde{X}}_{\beta}]= i\hbar_e \tilde{J}_{\alpha\beta}= - (\tilde{\Sigma}_y)_{\alpha\beta},
\end{equation}
where the deformed symplectic matrix $\tilde{J}$ and the effective Planck constant $\hbar_e$ reads
\begin{eqnarray}
\hat{\tilde{J}} = \left(\begin{array}{cc}
\hat{J}_2 &  \frac{1}{\hbar_e}\hat{\Pi}_{\theta\eta}\\
	-\frac{1}{\hbar_e}\hat{\Pi}_{\theta\eta} &  \hat{J}_2
\end{array}\right),\; \mbox{with}\; 
\hat{\Pi}_{\theta\eta}= \left(\begin{array}{cc}
	\theta &  0\\
	0 & \eta
\end{array}\right),\; 
\hbar_{e}=\hbar(1+\frac{\theta\eta}{4\hbar^2}).
\end{eqnarray}
Here $\theta$ and $\eta$ are the position position NC-parameter and momentum-momentum NC-parameter, respectively.
The NC-space co-ordinates ($\tilde{X}$) are connected to the  commutative space co-ordinates ($X$) through the Darboux transformation ($\hat{\Upsilon}_D$) given by the Bopp's shift
\begin{eqnarray}\label{cncconnection}
	\tilde{X}=\hat{\Upsilon}_D X,\;\;
\mbox{with}\;
	\hat{\Upsilon}_D =\left(\begin{array}{cc}
		\hat{	\mathbb{I}}_2 & -\frac{1}{2\hbar}\hat{\Pi}_{\theta\eta}\hat{J}_2 \\
		\frac{1}{2\hbar}\hat{\Pi}_{\theta\eta}\hat{J}_2 & \hat{\mathbb{I}}_2 
	\end{array}\right).
\end{eqnarray}
The notation $\hat{\mathbb{I}}_n$ stands for $n\times n$ identity matrix.
From the physical ground, the position-position and momentum momentum non-commutativity appear in a much higher energy scale. In other words $\theta<\hbar$ and $\eta< \hbar$. Hence the determinant of $\hat{\Upsilon}_D $ is nonzero ($\Delta_{\Upsilon_D}\neq 0$); i.e., $\hat{\Upsilon}_D \in GL(4, R)$.
$\hat{J}$ is connected with the deformed symplectic matrix $\hat{\tilde{J}}$ through 
\begin{equation}\label{JncJcconnection}
	\hbar_e \hat{\tilde{J}} = \hbar \hat{\Upsilon}_D \hat{J}\hat{\Upsilon}_D^T .
\end{equation}
Since the quantum mechanical formalism are well established in commutative space, it is customary to convert the NC-space system in the usual commutative space system through ~\eqref{cncconnection} for computational purpose.
The NC-space Hamiltonian for an anisotropic oscillator reads
\begin{equation}\label{NCShamiltonian}
	\hat{H}_{nc}=\frac{1}{2m_1}\hat{\tilde{P}}_1^2 + \frac{1}{2m_2}\hat{\tilde{P}}_2^2 + \frac{1}{2}m_1 \tilde{\omega}_1^2 \hat{\tilde{X}}_1^2 + \frac{1}{2}m_2 \tilde{\omega}_2^2\hat{\tilde{X}}_2^2 ,
\end{equation}
Using ~\eqref{cncconnection}, one can see that the NC-space Hamiltonian ~\eqref{NCShamiltonian}, corresponding to an anisotropic oscillator in NC-space is equivalent to the following commutative space Hamiltonian of an anisotropic charged oscillator in a magnetic field.
\begin{eqnarray}\label{Hamiltonianmatrix}
	\hat{H}=\frac{1}{2}p^\dagger\mu p + \frac{1}{2}x^\dagger Kx + x^\dagger L_1 p + p^\dagger L_2x ,\\
	\mbox{where}\; x= (x_1,x_2)^T,\; p=(p_1,p_2)^T.
\end{eqnarray}
Hermiticity of $\hat{H}$ implies $L_2=L_1^\dagger$. The explicit form of the effective mass matrix $\mu$, frequency matrix $K$ and angular momentum contributions $L_j$ are given by
\begin{eqnarray}
	\mu=\mbox{diag}(1/\mu_1,1/\mu_2),\; K=\mbox{diag}(\alpha_1,\alpha_2), \; L_1= \left(\begin{array}{cc}
		0 & -\nu_2 \\
		\nu_1 & 0
	\end{array}\right).
\end{eqnarray}
Explicit forms of the time-dependent parameters are given by
\begin{eqnarray}
	\frac{1}{\mu_1} &=& \frac{1}{m_1}+ \frac{1}{4\hbar^2} m_2\tilde{\omega}_2^2\theta^2, \;
	\frac{1}{\mu_2} = \frac{1}{m_2}+ \frac{1}{4\hbar^2} m_1\tilde{\omega}_1^2\theta^2,\\
	\alpha_1&=& \mu_1\omega_1^2=m_1\tilde{\omega}_1^2 + \frac{\eta^2}{4\hbar^2 m_2} ,\;
	\alpha_2  = \mu_2\omega_2^2=m_2\tilde{\omega}_2^2 + \frac{\eta^2}{4\hbar^2 m_1},\\
	\nu_1 & =& \frac{1}{4m_1\hbar} (\eta + m_1 m_2\tilde{\omega}_2^2 \theta),\;
	\nu_2 = \frac{1}{4m_2\hbar} (\eta + m_1 m_2\tilde{\omega}_1^2 \theta).
\end{eqnarray}
Note that for isotropic oscillator $\mu_1=\mu_2,\; \alpha_1=\alpha_2,\; \nu_1=\nu_2$. For an isotropic oscillator, one can identify $ x^\dagger L_1 p + p^\dagger L_2x $ as the angular momentum operator. 
\section{Time-dependent isotropic oscillator in NC-space} 
To make our study fairly general, we consider the parameters $\mu_i(t),\alpha_i(t),\nu_i(t)$ to be time-dependent (TD). We employ the Lewis-Riesenfeld (LR) method to solve TD-Schr\"{o}dinger equation corresponding to our TD-Hamiltonian.  LR-method  states that for a system described by a TD Hamiltonian $\hat{H} (t)$, a particular solution of the associated Schr\"{o}dinger equation is given by the eigenstate $ \vert n,t \rangle $ of a TD-invariant operator $ \hat{\mathcal{I}} $ defined by 
\begin{equation}\label{LRIcondition}
\partial_t\hat{\mathcal{I}} -i\hbar^{-1} [\hat{\mathcal{I}}, \hat{H}] = 0,
\end{equation}
 apart from a TD-phase factor $ e^{i \theta_n(t)} $, where $\theta_n(t)$ is given by
 \begin{equation}
\theta_n(t) = \int_{0}^t \langle n,\tau\vert (i\partial_t -\hat{H}(t))\vert n,\tau\rangle  d\tau.
 \end{equation} 
We illustrate this fact with an explicit closed-form solution for our system.
We consider the following ansatz for an invariant operator $\mathcal{I}$ corresponding to  ~\eqref{Hamiltonianmatrix}.
\begin{equation}\label{invariantansatzmatrix}
	\hat{\mathcal{I}}=p^\dagger A p + x^\dagger  B x + x^\dagger C p + p^\dagger D x .
\end{equation}
Hermiticity of $\hat{\mathcal{I}}$ implies
\begin{equation}
	A^\dagger =A,\; B^\dagger=B,\; D=C^\dagger.
\end{equation}
 Using the ansatz ~\eqref{invariantansatzmatrix} in ~\eqref{LRIcondition}, we obtain the following equivalent set of equations.
\begin{eqnarray}
	\dot{A} &=& (AL_1+L_2A)-(\mu C+C^\dagger \mu),\\
	\dot{B} &=& -(L_1B+BL_2)+(CK+KC^\dagger),\\
	\dot{C} &=& KA -B\mu + \left[C,L_1 \right],\\
	\dot{D} &=& AK-\mu B - \left[ D, L_2\right].
\end{eqnarray} 
 For an isotropic oscillator in noncommutative (NC) space, one can identify the Hamiltonian ~\eqref{Hamiltonianmatrix} with the following parameter values.
\begin{eqnarray}
	\mu= \mu_0^{-1}\mathbb{I}_2,\; K=\alpha \mathbb{I}_2,\; 	L_1=L_2^\dagger = -i\nu \sigma_y,\\
	\mbox{where}\; \frac{1}{\mu_0}=\frac{1}{m} + \frac{k\theta^2}{4\hbar^2},\; \alpha=k+\frac{\eta^2}{4m\hbar^2},\;
	 \nu = \frac{\eta}{4m\hbar} + \frac{k\theta}{4\hbar}.
\end{eqnarray}
What follows is that the invariance condition ~\eqref{LRIcondition} provides the following consistency condition for the invariant operator ~\eqref{invariantansatzmatrix}.
\begin{eqnarray}
	\dot{A} &=& (AL_1+L_2A)-(\mu C+C^\dagger \mu),\\
	\dot{B} &=& -(L_1B+BL_2) + CK +KC^\dagger,\\
	\dot{C} &=& KA-B\mu +\left[C,L_1\right],
\end{eqnarray}
where $A$ and $B$ are symmetric matrix.
The matrix $A,B,C$ satisfy the constraint equation
\begin{equation}\label{invarianceofisoncs}
	\frac{d}{dt} C_\kappa=i\nu \left[\sigma_y, C_\kappa\right],\;
\mbox{with}\;
	C_s=\frac{1}{2}(C+C^\dagger),\; C_\kappa = C_s^2-\frac{1}{2}\{A,B\}.
\end{equation}
For our present purpose, we choose a simple choice of invariant operator. This can be done by the observation of the following quasi-subalgebra.
\begin{eqnarray}
	\left[p^2,H_{incs}\right] &=& -i\hbar\alpha \mathcal{A}_{xp},\;
	\left[x^2,H_{incs}\right] = \frac{i\hbar}{\mu_0}\mathcal{A}_{xp},\\
	\left[\mathcal{A}_{xp},H_{incs}\right] &=& \frac{2i\hbar}{\mu_0}p^2  -2i\hbar\alpha x^2,\;
	\left[L,H\right] = 0,
\end{eqnarray}
where we have used the notations
\begin{equation}\label{angularmomentumopdefn}
	x^2=x_1^2+x_2^2,\; p^2=p_1^2+p_2^2,\; L=x_2p_1-x_1p_2,\; \mathcal{A}_{xp}=\{x_1,p_1\}+\{x_2,p_2\}.
\end{equation}
That means, one can choose a class of invariant operators of the form
\begin{equation}
	\mathcal{I}_{incs}= a_{11}p^2 + b_{11}x^2 +c_{11}\mathcal{A}_{xp} + 2lL.
\end{equation}
In that case we get  consistency conditions
\begin{eqnarray}\label{consistencyisoncs}
	\dot{a}_{11}=-2\mu_0^{-1}c_{11},\; \dot{b}_{11}=2\alpha c_{11},\; \dot{c}_{11}=\alpha a_{11} -\mu_0^{-1}b_{11},\; \dot{l}=0,
\end{eqnarray}
with the constraint
\begin{equation}\label{constraintisoncs}
	c_{11}^2- a_{11}b_{11}=-\kappa^2 =\mbox{constant}.
\end{equation}
  ~\eqref{consistencyisoncs} and ~\eqref{constraintisoncs} may be reduced to 
 the following integrable Ermakov-Pinney equation
\begin{equation}
	\ddot{\sigma}_{inc}+\frac{\dot{\mu}_0}{\mu_0}\dot{\sigma}_{inc}+\frac{\alpha}{\mu_0} \sigma_{inc} = \frac{\kappa^2}{\mu_0^2 \sigma_{inc}^3},\; \mbox{with}\; a_{11}(t)=\sigma_{inc}^2(t).
\end{equation}
For a diagonal representation of $\mathcal{I}_{incs}$, let us  write it in the quadratic form
\begin{eqnarray}
	\mathcal{I}_{incs} &=& X^\dagger \mathcal{H}_{incs}X,\\
\mbox{where}\;
	\mathcal{H}_{incs}&=& \left(\begin{array}{cc}
		M & L_{incs}^\dagger\\
		L_{incs} & M
	\end{array}\right),\; \mbox{with} \;
	M=\left(\begin{array}{cc}
		b_{11} & c_{11}\\
		c_{11} & a_{11}
	\end{array}\right),\;
	L_{incs}= il\sigma_y.
\end{eqnarray}
We wish to find a normal coordinate in which $\mathcal{I}_{incs}$ is diagonalized, keeping the intrinsic symplectic structure ($Sp(4,\mathbb{R})$) intact.  
The symplectic eigenvalues of $\mathcal{H}_{incs}$, are  the ordinary eigenvalues of the matrix $\Omega_{incs}=J\mathcal{H}_{incs}$, which has the
characteristic polynomial 
\begin{equation}
	P_\lambda=\lambda^4+2\tau\lambda^2+\Delta,\; 
\mbox{where}\;	\tau=l^2+\kappa^2,\; \Delta= (l^2-\kappa^2)^2.
\end{equation}
Hence $\Omega_{incs}$ has four purely imaginary eigenvalues
\begin{equation}\label{lambdaform}
	\lambda=\{-i\lambda_+,i\lambda_+,-i\lambda_-,i\lambda_-\}= \{-i\lambda_1,i\lambda_1,-i\lambda_2,i\lambda_2\},\; \mbox{where}\; \lambda_\pm= \kappa \pm l.
\end{equation}
Left eigenvector $u_{jl}=k_j(u_{j1},u_{j2},u_{j3},u_{j4})$ of $\Omega_{incs}$, corresponding to the eigenvalue $-i\lambda_j$, (i.e, $	u_{jl}\Omega_{incs}=-i\lambda_ju_{jl},\; j=1,2.$) may be written as
\begin{equation}
	u_{j\alpha}=u_{j\alpha r}+iu_{j\alpha c},\; \alpha=1,2,3,4;\; j=1,2.
\end{equation}
Here $k_j$ is the normalization constant. One set of choice is the followings.
\begin{eqnarray}
	u_{j1r} &=&u_{j3c}= 0,\;u_{j1c}= -2l\lambda_j b_{11},\;
	u_{j2r}= l(\lambda_j^2+\kappa^2-l^2),\; u_{j2c}=-2lc_{11}\lambda_j,\\
	u_{j3r} &=& b_{11}(\lambda_j^2-\kappa^2+l^2 ),\; 
	u_{j4r}= c_{11}(\lambda_j^2-\kappa^2+l^2),\; u_{j4c}=\lambda_j(\lambda_j^2 -\kappa^2-l^2).
\end{eqnarray}
The right-eigenvector $v_{jr}$ corresponding to the eigenvalue $-i\lambda_j$ is 
$v_{jr}=-\Sigma_y u_{jl}^\dagger$.
One can verify that, $\Omega_{incs}$ can be diagonalized with the following similarity transformation.
\begin{eqnarray}
\Omega_{incsD} &=&	Q_{incs}^{-1}\Omega_{incs} Q_{incs} = \mbox{diag}(-i\lambda_1,i\lambda_1,-i\lambda_2,i\lambda_2),\\
\mbox{where}\;	Q_{incs} &=& (v_{1r}, v_{1r}^*, v_{2r}, v_{2r}^*),\; Q_{incs}^{-1}=(u_{1l}^T, u_{1l}^{*T},u_{2l}^T,u_{2l}^{*T})^T.
\end{eqnarray}
Diagonal representation of $\Omega_{incs}$, enables us to  define the annihilation operators $\hat{a}_j$  and creation operators $\hat{a}_j^\dagger$ through
\begin{equation}
	\tilde{A}=(\tilde{a}_1,\tilde{a}_1^\dagger,\tilde{a}_2,\tilde{a}_2^\dagger)^T =Q_{incs}^{-1}X.
\end{equation}
Algebra of the annihilation and creation operators, i.e.,
\begin{equation}
	[\hat{\tilde{a}}_i,\hat{\tilde{a}}_j]=	[\hat{\tilde{a}}_i^\dagger,\hat{\tilde{a}}_j^\dagger]=0,\; 	[\hat{\tilde{a}}_i,\hat{\tilde{a}}_j^\dagger]=\delta_{ij};\; i,j=1,2,
\end{equation}
can be verified from the orthonormalization conditions of the left and right eigenvectors,
\begin{equation}
	u_iv_j=u_i^*v_j^*=\delta_{ij} ;\; i,j=1,2.
\end{equation} 
What follows is that we have obtained a factorization of $	\mathcal{I}_{incs}$ in terms of annihilation and creation operators. In particular,
\begin{eqnarray}
\mathcal{I}_{incs} &=&  -\tilde{A}^\dagger \Sigma_z Q_{incs}^{-1}\Sigma_y \mathcal{H}_{incs}Q_{incs}\tilde{A} \;\; (\because Q_{incs}^\dagger=-\Sigma_z Q_{incs}^{-1}\Sigma_y) \\
\therefore \mathcal{I}_{incs}  &=& i\tilde{A}^\dagger \Sigma_z \Omega_{incsD} \tilde{A} = \tilde{A}^\dagger \mbox{diag}( \lambda_1, \lambda_1, \lambda_2, \lambda_2 ) \tilde{A}
= 2\sum_{j=1}^{2} \lambda_j (\hat{\tilde{a}}_j^\dagger \hat{\tilde{a}}_j +1/2).
\end{eqnarray}
Eigenstates of $\mathcal{I}_{incs}$ may be expressed as
\begin{equation}
	\vert n_1,n_2\rangle_{incs} = \frac{1}{\sqrt{n_1!n_2!}} (\hat{\tilde{a}}_1^\dagger)^{n_1} (\hat{\tilde{a}}_2^\dagger)^{n_2} \vert 0,0\rangle_{incs} ,\;\; n_1,n_2=0,1,2,...,
\end{equation}
where  the ground state $\vert 0,0\rangle_{incs}$ of $\mathcal{I}_{incs}$ is obtained from
\begin{equation}\label{vacuumdefn}
	\hat{\tilde{a}}_1 \vert 0,0\rangle_{incs} = \hat{\tilde{a}}_2 \vert 0,0\rangle_{incs} = 0.
\end{equation}
In position representation ($\tilde{\psi}_{0,0}(x_1,x_2)=\langle x_1,x_2\vert 0,0\rangle_{incs}$), equation ~\eqref{vacuumdefn} reads
\begin{equation}\label{groundstatediffeqn}
	(	U_x x -i\hbar U_p \delta_x)\tilde{\psi}_{0,0}(x_1,x_2) = 0,
	\end{equation}
\begin{eqnarray}
\mbox{where}\;\; U_x = \left(\begin{array}{cc}
	u_{11} & u_{13}\\
	u_{21} & u_{23}
\end{array}\right),\;
U_p = \left(\begin{array}{cc}
	u_{12} & u_{14}\\
	u_{22} & u_{24}
\end{array}\right),\; 
	\delta_x=\left(\begin{array}{c}
		\frac{\partial}{\partial x_1}\\
		\frac{\partial}{\partial x_2}
	\end{array}\right).
\end{eqnarray}
To solve  equation ~\eqref{groundstatediffeqn}, we use the following ansatz. 
\begin{equation}
	\tilde{\psi}_{0,0}(x_1,x_2)= \mathcal{N}_0 e^{-\frac{1}{2} \sum_{i,j=1}^{2} x_i\Lambda^{s}_{ij}x_j},
\end{equation}
$\mathcal{N}_0$ being the normalization constant.
From the consistency condition, the explicit form of $\Lambda^{s}_{ij}$'s are given as follows.
\begin{eqnarray}
\tilde{\Lambda}^{s}=\frac{i}{\hbar}U_p^{-1}U_x,
	\mbox{where}\;\;
	\tilde{\Lambda}^{s}=\left(\begin{array}{cc}
		\Lambda_{11}^{s} & (\Lambda_{12}^{s}+\Lambda^{s}_{21})/2\\
		(\Lambda^{s}_{12}+\Lambda^{s}_{21})/2 & \Lambda^{s}_{22}
	\end{array}\right).
\end{eqnarray}
Explicitly written
\begin{eqnarray}
	\Lambda_{11}^{s} &=& 2\hbar^{-1}lb_{11} (\lambda_1-\lambda_2) [c_{11}(l^2-\kappa^2-\lambda_1\lambda_2)- i\lambda_1\lambda_2 (\lambda_1 +\lambda_2)]/\Delta_{U_p},\\
	\Lambda_{22}^{s} &=& 2\hbar^{-1}lb_{11} (\lambda_1-\lambda_2) [c_{11}(l^2-\kappa^2-\lambda_1\lambda_2)- i(\kappa^2 -l^2) (\lambda_1 +\lambda_2)]/\Delta_{U_p}.	
\end{eqnarray}
Where $\Delta_{U_p}$ is the determinant of $U_p$.
Using the explicit form ~\eqref{lambdaform}, we have
\begin{equation}
	\Delta_{U_p}= 8l^2(l^2-\kappa^2)(2c_{11}\kappa +i(\kappa^2-c_{11}^2)).
\end{equation}
And the matrix elements of $\tilde{\Lambda}$ reads
\begin{eqnarray}
	\Lambda_{11}^{s} = \Lambda_{22}^{s} &=& \Lambda_{11r}^{s} + i \Lambda_{11c}^{s} = \hbar^{-1} \sigma_{inc}^{-2} ( \kappa -i \mu_0 \sigma_{inc} \dot{\sigma}_{inc}),\\
	\Lambda_{12}^{s}+\Lambda_{21}^{s} &=& 0.
\end{eqnarray}
What follows is that the normalized ground state of $\mathcal{I}$ reads
\begin{equation}
	\tilde{\psi}_{0,0}(x_1,x_2) = \sqrt{\Lambda_{11r}^{s}/\pi} e^{-\frac{1}{2}\Lambda_{11}^{s}(x_1^2+x_2^2)}.
\end{equation}
According to Lewis-Riesenfeld theorem, the TD- Schr\"{o}dinger equation corresponding to $H_{incs}$ is satisfied by $	\tilde{\phi}_{0,0}(x_1,x_2,t) = \tilde{\psi}_{0,0}(x_1,x_2)e^{i\tilde{\theta}_{0,0}(t)}$
for some TD phase $\tilde{\theta}_{0,0}(t)$. Since the explicit form of $\tilde{\theta}_{0,0}(t)$ is not relevant for our present purpose, we keep it arbitrary.\\
What is crucial for our present discussion is that the state can be written as a product of states for two independent co-ordinates. In particular,
\begin{equation}
\tilde{\psi}_{0,0}(x_1,x_2)=\tilde{\psi}_{0}(x_1)\otimes \tilde{\psi}_{0}(x_2),\; \mbox{with}\; \tilde{\psi}_{0}(x_j)= \sqrt[4]{\Lambda_{11r}^s/\pi} e^{-\frac{1}{2}\Lambda_{11}^s x_j^2},
\end{equation}
which means the co-ordinate degrees of freedom $x_1$ and $x_2$ are not entangled. Deformation of symplectic structure with the help of noncommutative parameters can not ensure the entanglement. 
\section {Entangled states for time-independent anisotropic oscillator in NC-space}
We rewrite the Hamiltonian ~\eqref{Hamiltonianmatrix}, which is the commutative space equivalent for  an anisotropic oscillator in NC-space as
\begin{equation}\label{Hamiltonian}
	\hat{H}_{ancs}= \frac{1}{2\mu_1}\hat{p}_1^2 + \frac{1}{2\mu_2}\hat{p}_2^2 + \frac{1}{2}\mu_1\omega_1^2 \hat{x}_1^2 + \frac{1}{2}\mu_2\omega_2^2 \hat{x}_2^2 + \hat{L}_d,\;\mbox{with}\; 
	\hat{L}_d = \nu_1 \left\{\hat{x}_2,\hat{p}_1\right\} - \nu_2 \left\{ \hat{x}_1,\hat{p}_2  \right\}.
\end{equation}
Note that the deformed angular momentum operator $\hat{L}_d$ is reduced to the usual angular momentum operator $L$  ~\eqref{angularmomentumopdefn} for an isotropic oscillator. 
For an isotropic oscillator in NC-space, $\hat{L}$ commutes with the Hamiltonian.   However, for an anisotropic oscillator. ($\nu_1\neq \nu_2$), $\hat{L}_d$ does not commute with $H_{ancs}$ the process of diagonalization is not starightforward.
First, we write the Hamiltonian ~\eqref{Hamiltonian}  in the following quadratic form.
\begin{eqnarray}\label{quadraticH}
	\hat{H}_{ancs}=\frac{1}{2}X^T \hat{\mathcal{H}}_{ancs}X = 
	\frac{1}{2} X^T \left(  \begin{array}{cccc} \label{Hquadratic}
		\tilde{A}_I & \tilde{C}_I \\
		\tilde{C}_I^T & \tilde{B}_I
	\end{array}
	\right) X,
\end{eqnarray}
with
\begin{eqnarray}
	\tilde{C}_I = 2\left(\begin{array}{cc}
		0 & -\nu_2\\
		\nu_1 &0
	\end{array}\right),\;
	\tilde{A}_I=\mbox{diag}(\mu_1\omega_1^2,1/\mu_1),\; 
	\tilde{B}_I=\mbox{diag}(\mu_2\omega_2^2,1/\mu_2).
\end{eqnarray}
We wish to construct a normal coordinate system by diagonalizing $H_{ancs}$, maintaining the symplectic structure $Sp(4,\mathbb{R})$. The symplectic eigenvalues of $H_{ancs}$ are the ordinary eigenvalues of 
\begin{equation}
\Omega=JH_{ancs}=\left(\begin{array}{cc}
A_I & C_I \\
D_I & B_I
\end{array}\right) =
\left(\begin{array}{cc}
J_2\tilde{A}_I & J_2\tilde{C}_I \\
J_2\tilde{C}_I^T & J_2\tilde{B}_I
\end{array}\right).
\end{equation}
Since $\Omega^T\neq \Omega$, the left and right eigenvectors of $\Omega$ are not the same. However, left and right eigenvalues are identical. The characteristic polynomial $P_{\Omega}(\lambda)= \mbox{Det}(\Omega-\lambda\mathbb{I})$ of $\Omega$ is
\begin{eqnarray}
P_{\Omega}=\lambda^4 + \Delta \lambda^2 + \Delta_{\Omega},\; \mbox{with}\; \Delta=\Delta_{A_I} + \Delta_{B_I} +2 \Delta_{C_I}.
\end{eqnarray}
Here the notation $\Delta_{A}$ stands for $\Delta_{A}=\mbox{Det}(A)$. The discriminant
\begin{eqnarray}\label{discriminant}
	D^2=\Delta^2-4\Delta_\Omega = (\omega_1^2-\omega_2^2)^2 + 16\nu_1\nu_2 (\omega_1-\omega_2)^2 + 16 \left(\sqrt{\frac{\mu_1}{\mu_2}}\omega_1\nu_1 + \sqrt{\frac{\mu_2}{\mu_1}}\omega_2\nu_2\right)^2\ge 0.
\end{eqnarray}
Note that $D=0$ only for parameter values $\omega_1=\omega_2, nu_1=\nu_2=0$, which corresponds to the isotropic oscillator in commutative space. We shall consider $D>0$ for our present study.
Moreover, one can see
\begin{equation}
	\lambda^2= (-\Delta \pm D)/2 \le 0,
\end{equation}
which means, $	P_\Omega(\lambda)$ has four distinct purely imaginary roots for the  real parameters $\mu_j,\omega_j,\nu_j$. In particular,
\begin{equation}\label{lambdaformexplicit}
	\lambda \in \{\mp i\lambda_j, j=1,2\vert \lambda_1=\sqrt{(\Delta-D)/2}, \lambda_2=\sqrt{(\Delta+D)/2} \}.
\end{equation}
If $\chi_{lj}$ is the left eigenvector corresponding to the eigenvalue $-i\lambda_j$ of $\Omega$, i.e., $\chi_{lj}\Omega=-i\lambda_j \chi_{lj}$, then the direct computation gives
\begin{eqnarray}
\chi_{lj}=k_{lj}(x_{j1},x_{j2},x_{j3},x_{j4})=k_{lj}(i \kappa_{1j}, \kappa_{2j}, \kappa_{3j}, i\kappa_{4j});\; j=1,2.
\end{eqnarray}
Here $k_{lj}$ is the normalization constant, and the real parameters $\kappa_{ij} $ reads
\begin{eqnarray}
	\kappa_{1j} &=& -2\mu_1\lambda_j (\mu_1\nu_1\omega_1^2 + \mu_2\nu_2\omega_2^2),\;
	\kappa_{2j} = 2(\mu_2\nu_2\omega_2^2 - 4\mu_1\nu_1^2 \nu_2 + \mu_1\nu_1\lambda_j^2),\\
	\kappa_{3j} &=& \mu_1(4\mu_1\nu_1^2\omega_1^2 - \mu_2 \omega_1^2 \omega_2^2 + \mu_2\omega_2^2 \lambda_j^2),\;
	\kappa_{4j} = -\mu_1\lambda_j (\omega_1^2 + 4\nu_1\nu_2 -\lambda_j^2).
\end{eqnarray}
Left eigenvector corresponding to the eigenvalue $-i\lambda_j$ is given by $	\chi_{lj}^*$. Right eigenvector $\chi_{rj}$ corresponding to the eigenvalue $-i\lambda_j$   may be obtained through   $\chi_{rj}=-\Sigma_y \chi_{lj}^\dagger$. Normalization condition $\chi_{lj}\chi_{rj}=1$ yields
\begin{equation}\label{normalizationofchi}
	\vert k_{lj}\vert= 1/\sqrt{2(\kappa_{3j}\kappa_{4j}- \kappa_{1j} \kappa_{2j})};\; j=1,2.
\end{equation}
Similarity transformation which diagonalizes $\Omega$, i.e., 
\begin{equation}
	\Omega_D=\mbox{diag}(-i\lambda_1,i\lambda_1,-i\lambda_2,-i\lambda_2)= Q^{-1}\Omega Q,
\end{equation}
  is given by the following matrices
\begin{equation}
	Q=(\chi_{r1},\chi_{r1}^*,\chi_{r2},\chi_{r2}^*),\;	Q^{-1}=(\chi_{l1}^T,\chi_{l1}^{*T},\chi_{l2}^T,\chi_{l2}^{*T})^T.
\end{equation}
The diagonal representation of $\Omega$ enables to  define the normal co-ordinates through
\begin{equation}
	A=(a_1,a_1^\dagger,a_2,a_2^\dagger)^T=\frac{1}{\sqrt{\hbar}}Q^{-1}X.
\end{equation}
Using the normalization condition ~\eqref{normalizationofchi}, it follows 
\begin{equation}\label{algebraaadagger}
	[\hat{a}_1,\hat{a}_1^\dagger]=	[\hat{a}_2,\hat{a}_2^\dagger]=1.
\end{equation}
On the other hand, the orthogonality condition $\chi_{l1}\chi_{r2}=\chi_{l2}\chi_{r1}=0$ is equivalent to
\begin{equation}\label{algebraa1a2}
	[\hat{a}_1,\hat{a}_2]=0.
\end{equation}
The algebra ~\eqref{algebraaadagger} and ~\eqref{algebraa1a2} confirm that $\hat{a}_j$ and $\hat{a}_j^\dagger$ are annihilation and creation operators.
The ground state of the system thus satisfy the property
\begin{equation}\label{groundstatedefnforanisotropic}
	a_1\vert 0,0\rangle =a_2\vert 0,0\rangle =0.
\end{equation}
In position representation ($\psi_{0,0}(x_1,x_2)=\langle x_1,x_2\vert 0,0\rangle$), the equation ~\eqref{groundstatedefnforanisotropic} reads
\begin{equation}\label{groundstateforanisotropic}
	(U_xx-i\hbar U_p\partial_{x})\psi_{0,0}(x_1,x_2)=0,
\end{equation}
where
\begin{eqnarray}
	x=\left(\begin{array}{c}
		x_1 \\
		x_2
	\end{array}\right),\; 
	\partial_x=\left(\begin{array}{c}
		\frac{\partial}{\partial x_1} \\
		\frac{\partial}{\partial x_2}
	\end{array}\right),\;
	U_x=\left(\begin{array}{cc}
		i\kappa_{11} & \kappa_{31}\\
		i\kappa_{12} & \kappa_{32}
	\end{array}\right),\;
	U_p=\left(\begin{array}{cc}
	\kappa_{21} & i\kappa_{41}\\
	\kappa_{22} & i\kappa_{42}
	\end{array}\right).
\end{eqnarray}
We take the following ansatz for the solution of ~\eqref{groundstateforanisotropic}.
\begin{equation}\label{ansatzpsi00anisotropic}
	\psi_{0,0}(x_1,x_2)= \mathcal{N}_0 e^{-\frac{1}{2}x^T\Lambda_{ani}x}.
\end{equation}
$\mathcal{N}_0$ being the normalization condition. Using ~\eqref{ansatzpsi00anisotropic} in ~\eqref{groundstateforanisotropic} we get
\begin{equation}
	\tilde{\Lambda}_{s}=\frac{i}{\hbar} U_p^{-1}U_x,
\end{equation}
where
\begin{eqnarray}
	\Lambda_{ani}=\left(\begin{array}{cc}
		\Lambda_{11} & \Lambda_{12}\\
		\Lambda_{21} & \Lambda_{22}
	\end{array}\right),\;
	\tilde{\Lambda}_{s}=\left(\begin{array}{cc}
		\Lambda_{11} & (\Lambda_{12}+\Lambda_{21})/2\\
		(\Lambda_{12}+\Lambda_{21})/2 & \Lambda_{22}
	\end{array}\right).
\end{eqnarray}
Explitly written,
\begin{eqnarray}
	\Lambda_{11} &=& \frac{\kappa_{41}\kappa_{12}-\kappa_{42}\kappa_{11}}{\hbar(\kappa_{21}\kappa_{42}-\kappa_{22}\kappa_{41})} =\Lambda_{11r},\;\;
	\Lambda_{22} = \frac{\kappa_{21}\kappa_{32}-\kappa_{22}\kappa_{31}}{\hbar(\kappa_{21}\kappa_{42}-\kappa_{22}\kappa_{41})}=\Lambda_{22r}, \label{Lambda11form}\\
	\frac{1}{2}(	\Lambda_{12}+\Lambda_{21} ) &=& \frac{i(\kappa_{42}\kappa_{31}-\kappa_{41}\kappa_{32})}{\hbar(\kappa_{21}\kappa_{42}-\kappa_{22}\kappa_{41})}=\frac{i(\kappa_{12}\kappa_{21}-\kappa_{22}\kappa_{11})}{\hbar(\kappa_{21}\kappa_{42}-\kappa_{22}\kappa_{41})}\label{Lambda12form}=i\Lambda_{12c}.
\end{eqnarray}
In particular, $\Lambda_{11}$ and $\Lambda_{22}$ are real, whereas $\Lambda_{12}$ is purely imaginary. 
\section{Noise matrix and separability criterion}
Wigner distribution ($W$) and the density operator $(\hat{\rho})$ are related
through the definition
\begin{equation}
	W(X_c) =\frac{1}{\pi^2\hbar^2}\int d^2 x' \langle x-x'\vert \hat{\rho} \vert x+x'\rangle e^{2ix'.p/\hbar},
\end{equation}
where $X_c=(X_1,X_2,X_3,X_4)=(x_1,p_1,x_2,p_2)^T$ is the classical co-ordinate  vector. We define 
\begin{eqnarray}
	\Delta \hat{X}_\alpha &=& \hat{X}_\alpha -\langle \hat{X}_\alpha\rangle_\rho,  \mbox{with}\; \langle \hat{X}_\alpha\rangle_\rho = \mbox{Tr}(\hat{X}_\alpha\hat{\rho}); \;\alpha=1,2,3,4.\\
		\Delta X_\alpha &=& X_\alpha -\langle X_\alpha\rangle_W,\; \mbox{with}\; \langle X_\alpha\rangle_W = \int   W(X_c)X_\alpha d^4X_c .
\end{eqnarray}
The four components $\Delta \hat{X}$ satisfy the same commutation relations as $\hat{X}$. Moreover, the phase-space average $\langle X_\alpha\rangle_W$ with respect to the Wigner distribution $W$ is equal to the average $\langle \hat{X}_\alpha\rangle_\rho$ with respect to the density operator $\hat{\rho}$. We define the covariance matrix $\mathcal{V}_c$ through the matrix elements
\begin{equation}
	\mathcal{V}_{\alpha\beta}= \langle \{\Delta \hat{X}_\alpha, \Delta \hat{X}_\beta\}\rangle =\mbox{Tr}(\{\Delta \hat{X}_\alpha, \Delta \hat{X}_\beta\}\hat{\rho})=\int \Delta X_\alpha \Delta X_\beta W(X_c)d^4X_c,
\end{equation}
where we have used the notation $\{\Delta \hat{X}_\alpha, \Delta \hat{X}_\beta\}=(\Delta \hat{X}_\alpha \Delta \hat{X}_\beta + \Delta \hat{X}_\beta \Delta \hat{X}_\alpha)/2$.
It is evident that the covariance matrix $\mathcal{V}_c$ is symmetric. Hence, $\hat{\mathcal{V}}_c$ can be written in the block form
\begin{eqnarray}\label{cspacecovariancematrix}
	\hat{\mathcal{V}}_c=\left(\begin{array}{cc}
		V_{11} & V_{12}\\
		V_{12}^T & V_{22}
	\end{array}\right),
\end{eqnarray}
Using the connection ~\eqref{cncconnection} between the commutative space and NC-space co-ordinates, one can see that the NC-space covariance matrix is given by
\begin{equation}\label{VncVcconnection}
	\hat{\tilde{\mathcal{V}}}_{nc}= \hat{\Upsilon}_D  \hat{\mathcal{V}}_c\hat{\Upsilon}_D^T .
\end{equation}
The fundamental commutation relations ~\eqref{Cspacecommutation} of commutative space implies that a bonafide covariance matrix must satisfy the Robertson-Schr\"{o}dinger uncertainty principle (RSUP)
\begin{equation}\label{RSUPc}
	\hat{\mathcal{V}}_c+\frac{i}{2}\hbar\hat{J} \ge 0 .
\end{equation}
In NC-space, the equivalent statement for the RSUP reads
\begin{equation}\label{RSUPnc}
	\hat{\tilde{\mathcal{V}}}_{nc} + \frac{i}{2} \hbar_e \hat{\tilde{J}} \ge 0.
\end{equation}
A generic local transformation $\hat{S}_1 \bigoplus \hat{S}_2$, acts on $\hat{\mathcal{V}}_c$ as
\begin{equation}
	\hat{V}_{jj}\to \hat{S}_j \hat{V}_{jj}\hat{S}_j^T ,\; 	\hat{V}_{12}\to \hat{S}_1 \hat{V}_{12}\hat{S}_2^T;\; \mbox{with}\; \hat{S}_j \in Sp(2,\mathbb{R}),\;  j=1,2.
\end{equation}
One can identify that following four quantities are local  invariant with respect to transformation belonging to the $Sp(2,\mathbb{R})\bigotimes Sp(2,\mathbb{R}) \subset Sp(4,\mathbb{R})$.
\begin{eqnarray}
	\Delta_j=Det(V_{jj}),	\Delta_{12}=Det(V_{12}),  \Delta_{\mathcal{V}_c} = Det(\mathcal{V}_c),
	\tau_{v}=\mbox{Tr}(V_{11} J_2 V_{12} J_2 V_{22} J_2 V_{12}^T J_2).
\end{eqnarray}
Using Williamson's theorem \cite{Williamson1,Williamson2}, one can show that the RSUP ~\eqref{RSUPc} can be rewritten as $Sp(2,\mathbb{R})\bigotimes Sp(2,\mathbb{R})$ invariant statement
\begin{equation}\label{covariancersupc}
	\Delta_1 \Delta_2 + (\hbar^2/4- \Delta_{12})^2 -\tau_{v}\ge \hbar^2( \Delta_1+\Delta_2)/4.
\end{equation}
Under mirror reflection (Peres-Horodecki partial transpose) $\Delta_1$,  $\Delta_2$ and $\tau_v$ remains invariant; whereas $\Delta_{12}$ flips sign. Therefore, the requirement that the covariance matrix of a separable state has to obey the following necessary condition.
\begin{equation}\label{separabilityc}
	Ps=	\Delta_1 \Delta_2 + (\hbar^2/4-\vert \Delta_{12}\vert)^2 - \tau_{v}- \hbar^2( \Delta_1+\Delta_2)/4 \ge 0,
\end{equation}
which turns out to be sufficient for all bipartite-Gaussian state.
\section{Separability condition for time-independent anisotropic oscillator in NC-space }
For a generic ground state ~\eqref{groundstatedefnforanisotropic}, which has the form in position representation as
$\psi(x_1,x_2)=\mathcal{N}_0e^{-\frac{1}{2}x^T\Lambda x}$, corresponds to 
the Wigner distribution
\begin{equation}\label{WX}
W(X)= \frac{1}{\pi^2\hbar^2} \exp\{ - x^T (\Lambda_r + \Lambda_c \Lambda_r^{-1}\Lambda_c^T)x - \frac{1}{\hbar^2} p^T\Lambda_r^{-1}p  -\frac{1}{\hbar} (x^T \Lambda_c \Lambda_r^{-1} p + p^T \Lambda_r^{-1}\Lambda_c^T x) \}.
\end{equation}
Here the following notation for the matrix $\Lambda$ and its elements $\Lambda_{jk}$ have been used.
\begin{eqnarray}
	\Lambda=[\Lambda_{jk}]_{j,k=1}^{2}= \Lambda_r +i\Lambda_c= [\Lambda_{jkr}]_{j,k=1}^{2}+i[\Lambda_{jkc}]_{j,k=1}^{2},
\end{eqnarray}
 where $\Lambda_{jkr}=\Re(\Lambda_{jk}),\Lambda_{jkc}=\Im(\Lambda_{jk})$.
~\eqref{WX} can be written in the following compact form to illuminate its gaussian form.
\begin{equation}\label{wignerforanisoosc}
	W(x_1,x_2,p_1,p_2)= \frac{1}{\pi^2\hbar^2} e^{-\tilde{X}_c^T\tilde{\Lambda}_m\tilde{X}_c},
\end{equation}
with
\begin{eqnarray}
\tilde{X}_c=(x_1,x_2,p_1,p_2)^T = (X_{m1},X_{m2},X_{m3},X_{m4}),\;
	\tilde{\Lambda}_m = \left( \begin{array}{cc}
		\tilde{\Lambda}_1 & \tilde{\Lambda}_{12}\\
		\tilde{\Lambda}_{12}^T & \tilde{\Lambda}_2
	\end{array}\right),
\end{eqnarray}
where
\begin{equation}\label{LambdarLambdacform}
\tilde{\Lambda}_1=\Lambda_r + \Lambda_c \Lambda_r^{-1}\Lambda_c^T,\;
\tilde{\Lambda}_2 = \hbar^{-2} \Lambda_r^{-1},\;
\tilde{\Lambda}_{12}= \hbar^{-1}\Lambda_c\Lambda_r^{-1}.
\end{equation}
Using ~\eqref{wignerforanisoosc}, the  direct computation indicates
\begin{equation}
	\langle X_{m\alpha}\rangle_W=0 \implies 	\mathcal{V}_{\alpha\beta}=\int X_{m\alpha}  X_{m\beta} W(\tilde{X}_c)d^4\tilde{X}_c;\;\; \alpha,\beta=1,2,3,4.
\end{equation}
In particular, the covariance matrix is explicitly given by 
\begin{equation}
	\tilde{\mathcal{\sigma}}_{\mbox{cov}} = \frac{1}{2 \hbar^2 \sqrt{\Delta_{\tilde{\Lambda}_m}}} \tilde{\Lambda}_{m}^{-1}.
\end{equation}
Since $\tilde{\Lambda}_{m}$ is a partitioned symmetric matrix,  its inverse is also a partitioned symmetric matrix. In particular,
\begin{eqnarray}\label{Lambdaminverseform}
	\tilde{\Lambda}_{m}^{-1}= \left(\begin{array}{cc}
  (\tilde{\Lambda}_1 - \tilde{\Lambda}_{12} \tilde{\Lambda}_{2}^{-1}\tilde{\Lambda}_{12}^T)^{-1} &  - (\tilde{\Lambda}_2^{-1}\tilde{\Lambda}_{12}^T (\tilde{\Lambda}_1-\tilde{\Lambda}_{12} \tilde{\Lambda}_{2}^{-1} \tilde{\Lambda}_{12}^T)^{-1})^T\\
  -\tilde{\Lambda}_2^{-1}\tilde{\Lambda}_{12}^T (\tilde{\Lambda}_1-\tilde{\Lambda}_{12} \tilde{\Lambda}_{2}^{-1} \tilde{\Lambda}_{12}^T)^{-1} & \tilde{\Lambda}_2^{-1}+\tilde{\Lambda}_2^{-1} \tilde{\Lambda}_{12}^T ( \tilde{\Lambda}_1 - \tilde{\Lambda}_{12} \tilde{\Lambda}_2^{-1} \tilde{\Lambda}_{12}^T)^{-1} \tilde{\Lambda}_{12} \tilde{\Lambda}_2^{-1}
\end{array}\right)
\end{eqnarray}
Using ~\eqref{LambdarLambdacform}, we simplify ~\eqref{Lambdaminverseform} as
\begin{eqnarray}
\tilde{\Lambda}_{m}^{-1}= \hbar^2\left(\begin{array}{cc}
	\tilde{\Lambda}_2 & - \tilde{\Lambda}_{12}^T \\
	- \tilde{\Lambda}_{12} & \tilde{\Lambda}_1
\end{array}\right).
\end{eqnarray}
Moreover, the determinant of $\tilde{\Lambda}_{m}$ is given by
\begin{eqnarray}
\Delta_{\tilde{\Lambda}_m}= \mbox{Det}(\tilde{\Lambda}_2)\mbox{Det}(\tilde{\Lambda}_1 - \tilde{\Lambda}_{12} \tilde{\Lambda}_2^{-1}\tilde{\Lambda}_{12}^T) 
= \mbox{Det}(\hbar^{-2}\Lambda_r^{-1})\mbox{Det}(\Lambda_r)= 1/\hbar^4.
\end{eqnarray}
Note that, the covariance matrix $\tilde{\sigma}_{\mbox{cov}}$ corresponds to the co-ordinate ordering of $\tilde{X}_c$, which is related to our ordering of original co-ordinates $X_c$ as
\begin{eqnarray}
	X_c= \tilde{\mathcal{S}}_m\tilde{X}_c,\;\mbox{with}\; 
	\tilde{\mathcal{S}}_m =\left(\begin{array}{cc}
		\tilde{\mathcal{S}}_1 & \tilde{\mathcal{S}}_3\\
		\tilde{\mathcal{S}}_2 & \tilde{\mathcal{S}}_4
	\end{array}\right)
\end{eqnarray}
where
\begin{eqnarray}
\tilde{\mathcal{S}}_1 =\left(\begin{array}{cc}
1 & 0\\
	0 & 0
\end{array}\right),\;
\tilde{\mathcal{S}}_2 =\left(\begin{array}{cc}
	0 & 1\\
	0 & 0
\end{array}\right),\;
\tilde{\mathcal{S}}_3 =\left(\begin{array}{cc}
	0 & 0\\
	1 & 0
\end{array}\right),\;
\tilde{\mathcal{S}}_4 =\left(\begin{array}{cc}
	0 & 0\\
	0 & 1
\end{array}\right).
\end{eqnarray}
The covariance matrix $\mathcal{V}_{cov}$ corresponding to the ordering  $X_c$ is related to $\tilde{\sigma}_{\mbox{cov}}$ as
\begin{eqnarray}
	\mathcal{V}_{cov}= \left(\begin{array}{cc}
V_{11} & V_{12} \\
V_{12}^T & V_{22}
	\end{array}\right)=\tilde{\mathcal{S}}_m^T \tilde{\sigma}_{\mbox{cov}} \tilde{\mathcal{S}}_m = \frac{\hbar}{2}\left(\begin{array}{cc}
\sigma_{11} & \sigma_{12}\\
\sigma_{12}^T & \sigma_{22}
	\end{array}\right).
\end{eqnarray}
So in our notation $V_{ij}=\frac{\hbar}{2}\sigma_{ij}$.
The explicit form of  $\sigma_{ij}$ reads
\begin{eqnarray}
	\sigma_{11} &=& \hbar(\tilde{\mathcal{S}}_1 \tilde{\Lambda}_{2}\tilde{\mathcal{S}}_1 - \tilde{\mathcal{S}}_1 \tilde{\Lambda}_{12}^T \tilde{\mathcal{S}}_2 - \tilde{\mathcal{S}}_3 \tilde{\Lambda}_{12}\tilde{\mathcal{S}}_1 + \tilde{\mathcal{S}}_3 \tilde{\Lambda}_{1} \tilde{\mathcal{S}}_2), \\
	\sigma_{12} &=& \hbar(\tilde{\mathcal{S}}_1 \tilde{\Lambda}_{2}\tilde{\mathcal{S}}_3 - \tilde{\mathcal{S}}_1 \tilde{\Lambda}_{12}^T \tilde{\mathcal{S}}_4 - \tilde{\mathcal{S}}_3 \tilde{\Lambda}_{12}\tilde{\mathcal{S}}_3 + \tilde{\mathcal{S}}_3 \tilde{\Lambda}_{1} \tilde{\mathcal{S}}_4), \\
	\sigma_{22} &=& \hbar(\tilde{\mathcal{S}}_2 \tilde{\Lambda}_{2}\tilde{\mathcal{S}}_3 - \tilde{\mathcal{S}}_2 \tilde{\Lambda}_{12}^T \tilde{\mathcal{S}}_4 - \tilde{\mathcal{S}}_4 \tilde{\Lambda}_{12}\tilde{\mathcal{S}}_3 + \tilde{\mathcal{S}}_4 \tilde{\Lambda}_{1} \tilde{\mathcal{S}}_4). 
\end{eqnarray}
The matrix elements of $\sigma_{ij}$ are dimensionless. The explicit expressions ~\eqref{Lambda11form} and ~\eqref{Lambda12form} are
\begin{eqnarray}\label{formofsigmaij}
	\sigma_{11}=\left(\begin{array}{cc}
\frac{1}{\hbar\Lambda_{11r}} & 0\\
0 & \frac{\hbar\Delta_\Lambda}{\Lambda_{22r}}
	\end{array}\right),\;
	\sigma_{22}=\left(\begin{array}{cc}
	\frac{1}{\hbar\Lambda_{22r}} & 0\\
	0 & \frac{\hbar\Delta_\Lambda}{\Lambda_{11r}}
\end{array}\right),\;
	\sigma_{12}=\left(\begin{array}{cc}
0& -	\frac{\Lambda_{12c}}{\Lambda_{11r}} \\
-\frac{\Lambda_{12c}}{\Lambda_{22r}} &0
\end{array}\right),
\end{eqnarray}
where $\Delta_\Lambda= \Lambda_{11r}\Lambda_{22r}+\Lambda_{12c}^2$.
Using  ~\eqref{formofsigmaij} in the generalized Peres-Horodecki separability criterion ~\eqref{separabilityc} we get the following constraint on the parameters.
\begin{equation}
	-\Lambda_{11r} \Lambda_{22r} \Lambda_{12c}^2 \ge \Lambda_{12c}^2 \Lambda_{11r}\Lambda_{22r},
\end{equation}
which means
\begin{equation}
	\Lambda_{11r}\Lambda_{22r}\Lambda_{12c}=0.
\end{equation}
Now for nontrivial case $\Lambda_{12c}=0$, which implies
\begin{equation}\label{constrainteqn}
	(\lambda_2-\lambda_1)(\mu_2\nu_2\omega_2^2 -4\mu_1\nu_1^2\nu_2-\mu_1\nu_1\lambda_1\lambda_2)=0.
\end{equation}
However, according to ~\eqref{discriminant} and ~\eqref{lambdaformexplicit}, we consider $\lambda_1\neq \lambda_2$. ~\eqref{constrainteqn} then implies
\begin{equation}
(\mu_1^2\omega_1^2\nu_1^2-\mu_1^2\omega_1^2\nu_1^2)(\omega_2^2-4\mu_1\nu_1^2/\mu_2)=0.
\end{equation}
From the physical viewpoint, we choose $\theta,\eta \le \hbar$. In other words, $\theta\eta \neq 4\hbar^2$, which means $\omega_2^2 \neq 4\mu_1\nu_1^2/\mu_2$. Therefore, the only possibility for the separable states is satisfied by the constraint $\mu_1\nu_1\omega_1 =\mu_2\nu_2\omega_2$, which is equivalent to the following equation in terms of original parameters.
\begin{eqnarray}\label{sep1}
	(4\hbar^2/m_{12}+\tilde{\omega}_1^2\theta^2) (\eta/m_{12}+ \tilde{\omega}_2^2\theta)^2 (\eta^2/m_{12}+ 4\hbar^2\tilde{\omega}_1^2) 
	 =  (4\hbar^2/m_{12}+\tilde{\omega}_2^2\theta^2) \nonumber \\
	  (\eta/m_{12}+ \tilde{\omega}_1^2\theta)^2 (\eta^2/m_{12}+ 4\hbar^2\tilde{\omega}_2^2) ,\; \mbox{with}\; m_{12}=m_1m_2.
\end{eqnarray}
The condition ~\eqref{sep1} is trivially satisfied for commutative space ($\theta,\eta\to 0$), and as well as for isotropic oscillator ($\tilde{\omega}_1=\tilde{\omega}_2$) in noncommutative (NC) space. Therefore, the entanglement between the coordinate degrees of freedom is not the sole property of the noncommutativity of space. It depends on both the NC-parameters ($\theta,\eta$) and the anisotropy of oscillator frequency ($\tilde{\omega}_1\neq\tilde{\omega}_2$). Moreover, even in NC-space, an anisotropic oscillator also supports separable states. For instance, let us consider without loss of generality $m_1m_2=1, \hbar=1, \mu=1,\nu=1$, which provides the relation $\tilde{\omega}_2=1/\tilde{\omega}_1$ between the oscillator frequencies for which it supports separable states.
All the other frequency admits the entanglement between coordinate degrees of freedom. Since the separability of the bipartite Gaussian state for an anisotropic oscillator in NCS is satisfied for only a very special choice of anisotropic parameters, it is worth mentioning that the bipartite Gaussian state  is almost always entangled in NC-space.
\section{Conclusions}
We have shown that the mere inclusion of position-position and momentum-momentum noncommutativity is not sufficient to generate entanglement between coordinate degrees of freedom. Rather the entanglement property depends on both the anisotropy of oscillators and the noncommutative parameters. To make the present study fairly general, we incorporate the time-dependent parameters for isotropic oscillators. It turns out that the inclusion of arbitrary time dependence in all parameters is not sufficient to support entangled states. 
We have considered both the position-position and momentum-momentum noncommutativity under consideration. The Bopp's shift has been employed so that we can perform the equivalent computation in the usual commutative space. We have restricted our study to the bipartite Gaussian system mainly for the reason that the computational mechanisms are readily available for the same.\\
From the experimental perspective, if we observe the entanglement effects
in the measurement of the properties of two quantum masses that interact with each other
through gravity only, then we can conclude that the mediator (gravity) has some
quantum features \cite{marletto1,bose1}. It doesn't matter in what way gravity is quantum - whether it's loop quantum gravity or string theory or something else - but it has to be a quantum theory. In this respect, it is worth mentioning the proposed tabletop experimental setup of \cite{expt1,expt2,expt3,expt4,expt5} with the help of an optomechanical scheme to determine the NC-space parameter. In particular, the optomechanical scheme relies on the interaction with the high-intensity optical pulse in terms of a sequence of optomechanical interaction inside an optical resonator. The phase difference of the input and output signal provides a signature of the existence of NC parameters. The outcome of this present paper indicates that the oscillator inside the optical cavity has to be anisotropic to observe any measurable effect of entanglement between coordinate degrees of freedom.  \\
Since the covariance matrix can be used to construct all the relevant information for the bipartite system, one can construct the thermodynamic quantities, e.g., extractable work from Szilard's engine cycle. Accordingly, whether the extractable work from the bipartite system with the association of coordinates with interaction entities can explain the expansion of space might be an interesting topic for further study. 
\section{Data availability statement} The manuscript has no associated data.
\section*{Acknowledgments :}
Pinaki Patra is grateful to Science and Engineering Research Board, Department of Science and Technology, Government of India for the financial support with the project's reference no. EEQ/2023/000784. SN and MR are grateful to Brahmananda Keshab Chandra College for their hospitality during the carry-out of their present project work.

\end{document}